# Marked variability in modern-time gravitational data indicates a large secular increase in the mass of ponderable bodies

Valerică Raicu*
*Kochi Medical School, Nankoku 783-8505, Kochi, Japan*

In spite of two hundred years of considerable efforts directed towards improvement in the experimental techniques, gravitational measurements have provided unsettled results for Newton's gravitational constant G. Analysis of the published (over ~75 years) small-scale gravitational measurements, presented in this report, unveils a large secular increase in the gravitational force that reveals itself as a formal increase in the Newton's constant G at a rate $\dot{G}/G = (1.43 \pm 0.08) \times 10^{-5}$ year$^{-1}$. Since its interpretation as a true 'G dot per G' effect is excluded by laser and radar ranging to the Moon and the interior planets as well as by double-pulsar studies, which all put tight limits on any variation in G, this large secular effect appears to originate in a temporal increase of gravitational masses due to, e.g., capture of mass from a hypothetical cold dark matter halo of the Sun. By virtue of the equivalence principle, the secular mass increase leads naturally to modifications of the Newtonian dynamics in the Solar System, as it predicts a cosmic deceleration of moving bodies dependent upon their speed. In particular, it predicts for the Pioneer 10/11 spacecraft an anomalous, nearly constant acceleration of $(-11.0 \pm 1.8) \times 10^{-10}$ m s$^{-2}$, in agreement with the published experimental value of $(-8.7 \pm 0.9) \times 10^{-10}$ m s$^{-2}$. Dynamical effects of this kind have been also detected in the motion of other spacecraft and artificial satellites, but not in the motion of large and/or old objects in the Solar System, such as the planets and their natural satellites.

---

*Present address: 708-10 Grenoble Dr., Toronto, ON, M3C 1C6, Canada;
E-mail: vraicu@pop.med.kochi-ms.ac.jp; Fax: (01) 416-696 9398

## I. INTRODUCTION

Newton's inverse square law, $F = Gm_1m_2/r^2$ – a cornerstone of classical gravitation –, giving the attraction force between two masses $m_1$ and $m_2$ set a distance $r$ apart, stands isolated in the larger edifice of physics. Its situation is essentially reflected by the absence of any experimentally verified equation relating the gravitational constant G to other fundamental physical constants. This is so partly because large uncertainties in the determination of G [1, 2], unparalleled by those of any other physical constant, preclude an accurate check of any potentially correct theory. At times, such uncertainties made room for interpretations that spurred searches for new physics, or cast doubts [3, 4], which now appear as unfounded though [5, 6], on the validity of Newton's inverse square law at terrestrial scales or on Einstein's equivalence principle. To help explain a plethora of otherwise poorly understood phenomena, particularly at astronomical scales [7], suggestions have been made for modifications of gravity at low accelerations [8] and/or large distances [9]. A majority's view, however, is that our accounting of the matter in the universe is incomplete [10] and that, if given the right input (i.e., including all forms of matter, luminous as well as dark), the classical (and Einsteinian) gravity theory should be able to produce the right results. Recently [11], a new phenomenon has been discovered that adds to the constellation of gravity's enigmas: an anomalous radial deceleration experienced by the Pioneer 10/11 spacecraft, which is at odds with the results of radio and laser ranging to the inner planets and the moon, and has received no satisfactory explanation to date [12, 13].

In the quest for the true value of the gravitational constant and for the all-encompassing theory of gravity, a route seems to have been insufficiently explored: the possibility that the infamous instability of Newton's G constant may in fact be not purely random, but it may, at least in part, arise from some systematic variation of the gravitational force, which, if unaccounted for, may be erroneously interpreted as a change in G. In this report, an analysis of modern-times literature data on G is presented, from which a rate of variation for the gravitational force is extracted that formally appears as a large 'G dot per G' effect. The effect is significant and requires proper interpretation. A conceptual model is therefore proposed that allows for such a dramatic effect to occur while not contradicting the traditional gravity tests and the tight limits they put on any variation in G [5, 14-16]. Then, important modifications induced by this secular effect upon Newtonian dynamics are worked out in detail, and their predictions are compared to available experimental data [12], and placed in a wider context of observational physics at astronomical scales [17].

## II. DATA SELECTION AND ANALYSIS

Available G data from the twentieth century small-scale experiments carrying errors of 0.2% or less (which amounts to rounding up the third digit by ± 1 unit) were used as primary data source for this analysis. This error criterion was rather "esthetical" because those data carrying large standard errors would be anyway weighted out in the fitting process. Much of the initial guidance to early papers was offered by the invaluable reviewing work of G.T. Gillies [18, 19] and by the useful discussions in the papers of Cohen and Taylor [20] and Long [4]. Sustained effort has been directed towards gaining access to most of the original papers, or to other citing papers that gave more detailed account of a particular work. I have also conducted my own searches to identify new results, particularly those that have been published after Gillies' reviews. Overall, there has been published an impressive number of papers on the determination of the gravitational constant, including those dealing with technical issues. However, not all of the investigated reports did eventually contribute a G value to the analysis in this paper, either because of not complying with our imposed accuracy limit of 0.2% or because, in some cases, results of the same experiments were reported in several places as the data analysis progressed. A number of papers have only signaled systematics that went previously undetected in a particular experiment, or just gave more precise indication on the date at which an experiment was carried out. Also removed from this analysis were the results from marine and geophysical surveys, and mine measurements, mostly because they are more prone to large systematic errors, due to limited knowledge of densities, borders, etc., and partly because I believe it was important to keep with relatively small size scale. However, this exclusion only sacrificed a few very good papers based on geophysical measurements [21].

Error assignment was done as follows. For a vast majority of the literature G values, the errors were obtained from the reported statistical and evaluated systematic errors combined in quadrature. Where the outcomes of series of experiments were all reported in a single paper together with their date of measurement (notably, a paper by Karagioz and Izmailov [22]), in order to avoid an overweighting of the present results by the data from a single research group, I have calculated an annual mean and the standard deviation, the latter being added quadratically to the reported systematic error. From all of the papers analyzed I have obtained a total of 40 values for G, collected in Table I, which met the requirements mentioned above.

The dates at which the experiments were carried out and their corresponding uncertainties were based on information furnished by the authors (for 21 data sets) or, if such information was lacking, the dates were computed as follows: Where a paper was directly available, but no precise time-related information was provided (9 data sets), the time ± its uncertainty were taken as [(*received date* – 0.5) ± 0.5] *year*. For results published in less popular journals as well as for older reports that were not directly available to me and for which no time-related information was provided by the citing article (7 cases), the time ± error were computed as [(publication date – 1) ± 1] *year*. For three other older experiments, summarized by the rows 1, 2 and 4 of Table I, comprehensive information was available from the citing paper.

As seen from Table I, there is a wealth of data concentrated over the past decade. Most of these many researches have been apparently stimulated by some contradicting results published in the first half of the nineteen nineties, and especially by the surprisingly high G value [$(6.71540 \pm 0.000083) \times 10^{-11}$ $m^3\, kg^{-1}\, s^{-2}$] reported by Michaelis et al in 1995 [23] (see also Refs. [1, 19] for comparison and discussions). The analysis in the present work does not include Michaelis' value, which evidently falls far outside the bulk of the data in Table I. Support for this exclusion comes also from a recent paper by Quinn et al [24] unveiling important source of errors that may have affected Michaelis et al's work.

The remaining 39 data sets in Table I were fitted, by using a weighted least-squares algorithm, to the linear function:

$$G(t) = G_0^{th}\left[1 + \frac{\dot{G}(t)}{G_0^{th}}\left(t - t_0\right)\right], \tag{1}$$

where $t$ is the date of measurements (in calendar years) and $G_0^{th}$ is predicted G value at $t_0$ = 1927. The best-fit parameter values were $G_0^{th}$ = $(6.6671 \pm 0.0005) \times 10^{-11}$

m$^3$ kg$^{-1}$ s$^{-2}$ and $\dot{G}/G$ = (1.47 ± 0.12) × 10$^{-5}$ year$^{-1}$, the errors being from covariant matrices associated with the least-square analysis. (Note that the indices were dropped from the second parameter, for simplicity.) The normalized chi-square ($\chi^2/d_f$, where $d_f$ is the number of degrees of freedom) associated with the fit was 2.74, which indicated that either the linear increase was not a suitable model for the data in Table I or that there still are outliers among the data. Analysis of the normalized residuals (i.e., the deviation of the data points from the best-fit line, expressed in units of the assumed experimental error) supported the second possibility as it unveiled three clear outliers: the data from rows 8 (obtained by Sagitov et al) and 40 (by Quinn et al) of Table I, which were 3 and 4.7 standard errors, respectively, above the best-fit line, and the value from row 33 (by Luo et al) whose normalized residual was -5.8.

Having these three additional values removed, the fitting was carried out again and the $\chi^2/d_f$ has dramatically improved to become 1.013. This gave a probability P of 45 % that a value of $\chi^2/d_f$ larger than the one just obtained can occur by chance, which was much larger than the 5 % value usually taken as the threshold under which a theoretical model is not supported by the data. Put in other words, the near-unity value of the normalized chi-square indicated that the difference between the best-fit line and each individual data point was in this case much smaller than the assigned experimental uncertainties, which validated the choice of Eq. (1) as a model for the variation of gravitational data.

The fitting results for the last case were plotted together with the experimental data in Fig. 1. The best-fit parameters were $G_0^{th}$ = (6.6672 ± 0.0003) × 10$^{-11}$ m$^3$ kg$^{-1}$ s$^{-2}$ and $\dot{G}/G$ = (1.43 ± 0.08) × 10$^{-5}$ year$^{-1}$, which did not differ significantly from those obtained with only one G value excluded. This suggested that the excluded G values were in overall agreement with the bulk of the data and that only their assigned experimental errors have been probably too optimistically evaluated. There is a small possibility though that the position of those points on the plot could have indeed reflected some real details in the temporal evolution of the gravitational force.

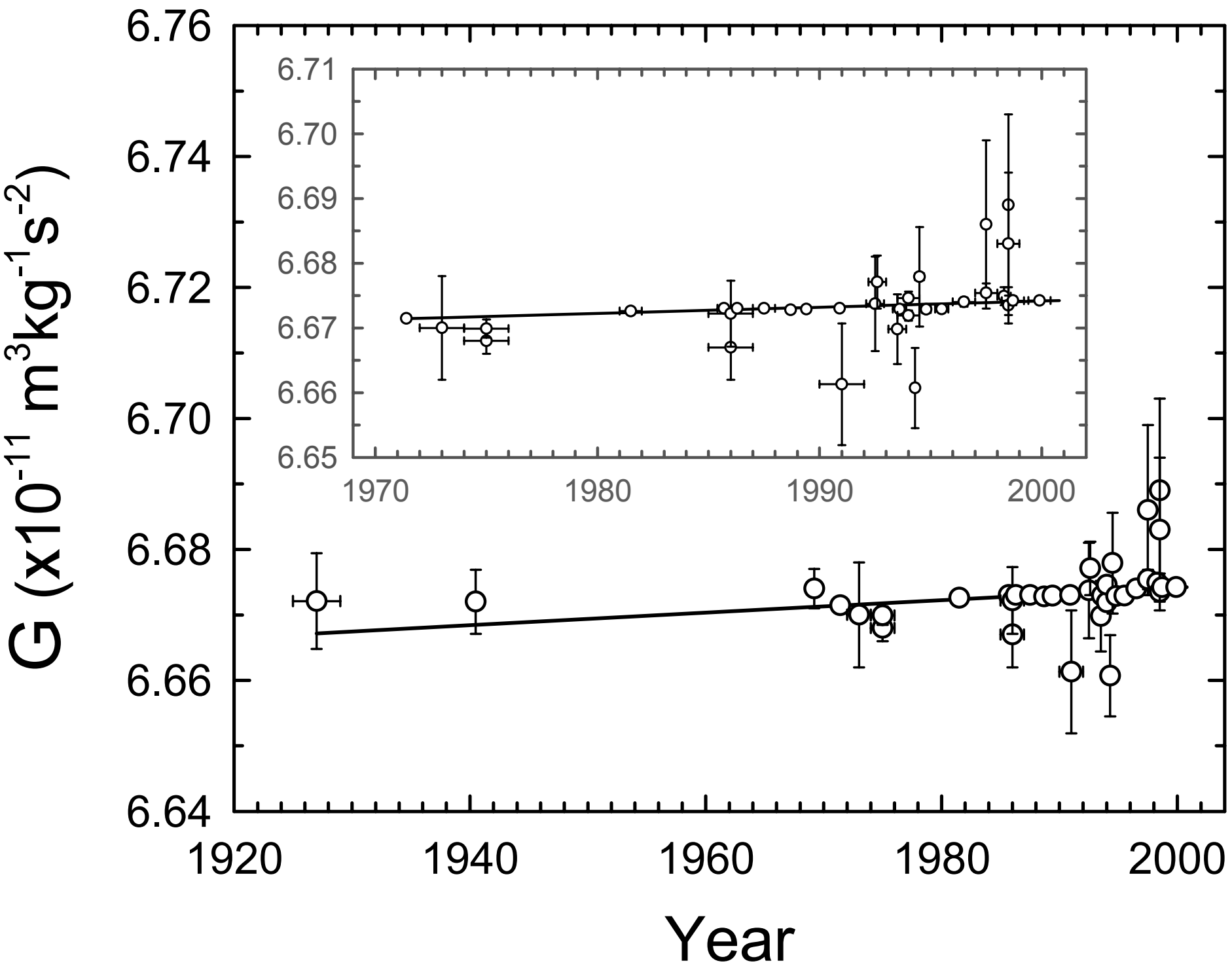

**FIG. 1.** Newton's gravitational constant G vs. time. Symbols represent the data from Table I, while the solid lines are best fits by Eq. (1). Inset, expanded plot for the last 30 years only.

The above analysis apparently indicated a very large secular increase in the gravitational force which, however, may not be interpreted as a true $\dot{G}/G$ effect, as it would exceed by five to seven orders of magnitude the limits on $\dot{G}/G$ inferred from planet and satellite ranging [15] and from studies of the binary pulsars [5, 16, 25]. If, on the other hand, this large systematic increase were a "mere" artefact, its isolation would be a difficult though necessary task to be undertaken by gravitation experimentalists.

There only remains for us here to explore the possibility that, while not actually reflecting a true increase in G, the observed secular increase in the gravitational force is the expression of some real physical phenomenon whose existence should not contradict a large body of knowledge, and which should also have other testable consequences.

## III. WAY OUT OF THE DILEMMA

Of the several possibilities of re-interpreting the observed increase in the gravitational force, a most productive one would be, as we shall see, to assume a secular increase of the masses involved in Newton's gravity law. The idea of mass and/or gravitational constant variation has been a recurring subject in physics, particularly since the proposal of the Dirac's large number hypothesis [26]. However, no theory so far has conceived such large variations in any of its parameters, as we have to consider here. Therefore, checking out how a large secular mass increase would affect various branches of science would be a very serious enterprise for anyone to undertake. Within the inherent size and scope constraints of this paper, however, I shall assume the more modest duty to envisage a scenario that leads to a parametric formula for the secular mass increase, whose rate (i.e., the model parameter) can be inferred from fitting the Newton's law to the gravitational data above, and whose validation is subsequently done by checking some of its implications for the dynamics of ponderable bodies against available experimental data.

As some people have already discussed in the wider context of dark matter in galaxies [7], our Sun may present a (spherical) halo of cold dark matter – constituted of e.g. heavy neutrinos – which could extend out to several light years [27, 28], and would make the Sun to behave as an extended particle. Experimental observations put limits of the order of $10^{-6}$ solar masses on the amount of dark matter contained within the orbits of outer planets [28, 29]. Inspired by such ideas, I developed a conceptual model based on the following postulates:

(1) The Sun carries with it a halo of (cold) dark matter extending with slowly decreasing (i.e., nearly constant) density to light years away from the Sun.

(2) An object moving through a cold dark matter halo captures, through impact with halo particles, small amounts of matter in direct proportion to its own mass and the distance d*x* traveled.

Based on these postulates, one can derive a formula for the secular mass increase of (extended) objects traveling through the Sun halo. To simplify maters and help emphasize the non-relativistic, yet, non-Newtonian character of the present phenomena, retardation effects due to noninstantaneous field propagation are ignored throughout this paper. A more general treatment will be attempted elsewhere.

According to the postulate 2, the amount of captured mass d*m* is given by:

$$\mathrm{d}m = \lambda m \mathrm{d}x = \lambda m v \mathrm{d}t\,, \tag{2}$$

where $v$ (= |**v**|) is body speed relative to background halo, and λ is a proportionality factor that we shall take as constant to simplify matters. Upon integrating the mass equation with respect to time, one obtains:

$$m(t) = m(t_0) e^{\lambda \int v dt}\,. \tag{3}$$

In terms of the mean speed $v_m$ over the time interval Δ*t*, Eq. (3) reads:

$$m(t) = m(t_0) e^{\lambda v_m \Delta t}\,. \tag{4}$$

Obviously, any object on the Earth surface moves together with it through the Sun's halo, and according to Eq. (4) it should experience a secular mass increase at a rate proportional to the Earth speed. Thus, if one tries to measure the gravitational attraction between two small objects placed on the Earth, one will detect a temporal variation in the gravitational attraction between the two bodies following the secular mass increase. According to Eq. (4), the force increase is given by a factor $e^{2\lambda v_m \Delta t}$ ("2" comes from the *two* masses) that reduces to $1 + 2\lambda v_m\,\Delta t$ in the limit of short time intervals (~$10^3$ years). If this is taken as the real cause for the apparent variation of G, and using the above value of $(1.43 \pm 0.08) \times 10^{-5}$ year$^{-1}$ for $\dot{G}/G$ and a mean speed of 29.8 ± 1.0 km/s for objects resting on the Earth surface (the error takes into account both the orbital and the rotational speed of the Earth), a value of $(7.61 \pm 0.67) \times 10^{-18}$ m$^{-1}$ is obtained for the constant λ. It is this short time limit value that will be considered in the discussions hereafter, because the time span of the available gravitational data is rather limited.

Another possible limitation of the model comes from its constant λ being calibrated against data from small-scale gravitational experiments, as mentioned above. Therefore, some or all of the model predictions may not be expected to directly apply to the behavior of very large bodies, such as planets and their natural satellites. A discussion on the case of planetary-size objects is made at the end of section IV.

## IV. SOME CONSEQUENCES OF THE SECULAR MASS EFFECT

### A. Principles of dynamics revisited

By virtue of the principle of equivalence between gravitational and inertial masses that should apply not only to ordinary matter but also to dark matter [30], the most readily predictable consequence of the secular increase in the gravitational mass is a modification of classical mechanics principles.

A force **F** acting on a body induces temporal changes in the body's momentum *m***v**; the latter may in turn be used as the definition of force. With the aid of Eq. (2) one may get, in usual notations:

$$\mathbf{F} = \frac{\mathrm{d}(m\,\mathbf{v})}{\mathrm{d}t} = m\dot{\mathbf{v}} + m\lambda v\mathbf{v}\,. \tag{5}$$

This equation differs from *Newton's second law* due to the time dependence of the mass.

If no external force, in the classical sense, acts upon the moving object, the motion equation becomes:

$$\dot{\mathbf{v}} = -\lambda v \mathbf{v} \tag{6}$$

which implies that "freely" moving bodies undergo a *cosmic deceleration* so that, in the limit of short time intervals, their speed varies as:

$$v \cong v(t_0) - \lambda v^2(t_0)\Delta t\,.$$

This is in stark contrast with *Newton's first law* that asserts that moving bodies should maintain indefinitely their uniform motion in the absence of external forces.

A similarity between this "cosmic" deceleration, and the anomalous acceleration experienced by the Pioneer 10/11 spacecraft [11, 12] catches our attention. However, the Pioneer problem also involves the solar gravitation, and we will discuss it not until laying down the equation of motion in gravitational field in the next sub-section.

Unlike the first two principles, *Newton's third law*, maintaining that the actions of two bodies upon each other are equal and directed oppositely, is left unchanged within the framework of this paper.

### B. Dynamics of gravitating bodies

The secular mass increase may in general result in a modification of the equation of motion for gravitating bodies. To illustrate, consider the simpler but convenient problem of a small body of mass *m* orbiting a central body (e.g., the Sun) of mass $M \gg m$ at a distance *r* in the body *M*'s proper frame. To calculate the force of attraction between the two bodies, we first write down the expression of the gravitational potential at a distance *r* from the Sun within the halo:

$$\Phi = -\frac{GM}{r} + 2\pi G\rho_H\left(\frac{1}{3}r^2 - r_H{}^2\right)$$

where the second term represents the potential within a spherically symmetrical distribution of mass of radius $r_H$ [7] and approximately constant density $\rho_H$. Because of this term, a small (negative) contribution is added to the Newtonian and general relativistic perihelion shift per century, as expressed by [27]:

$$\delta\omega = -\frac{4\pi^2\rho_H a^3}{M}\left(1-e^2\right)^{1/2}\cdot\frac{100}{T}, \tag{7}$$

where *a* is the semi-major axis, *e* is eccentricity of the unperturbed Keplerian orbit, and *T* is the unperturbed period expressed in years. In the case of the Icarus asteroid, whose orbital characteristics are used on occasions to test predictions of gravitation theories, this additional contribution to the perihelion precession, calculated from $\rho_H \leq 1 \times 10^{-14}$ kg m$^{-3}$ (as obtained from the upper limit for the dark matter mass within Uranus' orbit [29]), $M = 1.99 \times 10^{30}$ kg, $a = 1.08$ a.u., $e = 0.827$, and $T = 1.12$ y, is $\delta\omega \geq -0.009''$ per century – which is clearly undetectable at the present (the lower limit is $-1.3''$ per century for Icarus). Therefore, the second term in Eq. (7) makes no significant contribution to the perihelion precession of objects in the Solar System.

Let us next define the gravitational force experienced by the small body as:

$$\mathbf{F}_G = mG\nabla\left[\frac{M}{r} - 2\pi\rho_H\left(\frac{1}{3}r^2 - r_H{}^2\right)\right].$$

If $\rho_H$ varies in time due to capture of dark matter by the small body it has also a nonzero space gradient [31]. However this contribution to the total force can be safely ignored for nonrelativistic velocities. Besides, it is very likely that the halo regions beyond the small body orbit could readily redistribute to compensate the density decrease. In any event, we can get, to a good approximation, the following equation for the gravitational force:

$$\mathbf{F}_G = -\frac{GmM}{r^3}\mathbf{r} - \frac{4\pi}{3}Gm\rho_H\mathbf{r}. \qquad (8)$$

It has been shown above that the dark matter halo makes no presently detectable contribution to the perihelion shift of Icarus. Obviously, the second term in Eq. (8) too should vanish for any object whose orbit is interior to Uranus orbit, since the upper limit for the density $\rho_H$ has been determined [29] by using precisely this assumption that any perturbation is smaller than the detection limit. Nevertheless, we cannot discard that term right away, because the next sub-section deals with the dynamics of objects – the Pioneer 10/11 spacecraft – that are presently located far beyond all known planets of our Solar System, and it is not apparent that the halo-related term vanishes in their case.

By combining Eqs. (5) and (8) one gets the equation of motion of the small body,

$$\ddot{\mathbf{r}} + \lambda v\dot{\mathbf{r}} + \frac{GM}{r^3}\mathbf{r} + \frac{4\pi}{3}G\rho_H\mathbf{r} = 0, \qquad (9)$$

which departs markedly from the classical Kepler problem, mostly due to the viscous-drag-like term $\lambda v\dot{\mathbf{r}}$. I will discuss the implications of this term for the dynamics of small bodies in the next two subsections. To anticipate, however, it can be mentioned here that unlike the halo-related term in the gravitational potential, $\lambda v\dot{\mathbf{r}}$ produces no net perihelion shift over one complete revolution of a test body moving in a bound orbit [see Eq. (14) below].

The radial component of Eq. (9) is:

$$\ddot{r} + \lambda\left(\dot{r}^2 + r^2\dot{\theta}^2\right)^{1/2}\dot{r} - r\dot{\theta}^2 + \frac{GM}{r^2} + \frac{4\pi}{3}G\rho_H r = 0. \qquad (10)$$

while the angular one is:

$$\dot{l} + \lambda\left(\dot{r}^2 + r^2\dot{\theta}^2\right)^{1/2} l = 0 \qquad (11)$$

where $l = r^2\dot{\theta}$ is the specific angular momentum. The latter equation states that the angular momentum is dissipated at a rate dependent on velocity, which may cause a test body to fall towards the Sun. This effect will be discussed in Section IV. D.

### C. Radial motion and the "Pioneer" effect

Analysis of radio Doppler and ranging data from distant spacecraft (Pioneer 10/11) departing the solar system, elegantly done by Anderson and coworkers and reported in recent papers [11, 12], seems to support our inferred "cosmic" deceleration of moving bodies. Indeed, the two Pioneer spacecraft underwent an anomalous constant (over a period of ~10 years) radial deceleration with respect to the Sun, which could not be unequivocally explained [12, 13], either "prosaically" (i.e., by technological imperfections of the space travel and by known physics) or by various new as well as old theories.

Equation (10) provides an explanation: The Pioneers moving on hyperbolic orbits generate large radial speed components $\dot{r}$ of 12.0 ± 0.3 km/s, but negligible circular speeds $r\dot{\theta}$ (see Fig. 2). Thus, a radial acceleration,

$$\ddot{r} = -\lambda\dot{r}^2 - \frac{4\pi}{3}G\rho_H r - \frac{GM}{r^2}, \qquad (12)$$

is predicted, which departs from the classical mechanics' situation mainly by an 'anomalous' term $\lambda v\dot{\mathbf{r}}$. The absolute value of the second term in Eq. (12) is smaller than $0.3 \times 10^{-10}$ m s$^{-2}$ for a distance $r = 70$ a.u. that corresponds to the farthest position of the Pioneer 10 spacecraft in the Anderson et al study [12]. In fact, most of the time this term would have been even smaller, due to its dependence on distance. Furthermore, it is quite possible that the dark matter density in the Sun's halo is orders of magnitude lower than the upper limit determined from planetary dynamics, and comparable to the mean galactic density for nonluminous matter. In any case, the value obtained for the second term is small enough for it to remain unnoticed in the Pioneer spacecraft dynamics (see below); I will only add it to the acceleration assigned uncertainty. The Pioneer spacecraft anomalous acceleration is thus represented to a very good approximation by the equation:

$$\ddot{r}_P \cong -\lambda \dot{r}^2 , \qquad (13)$$

which, by taking $\lambda = (7.61 \pm 0.67) \times 10^{-18}$ m$^{-1}$ determined in section III, gives $(-11.0 \pm 1.5) \times 10^{-10}$ m s$^{-2}$ for the Pioneer residual acceleration. The final value obtained by considering a supplementary error of $0.3 \times 10^{-10}$ m s$^{-2}$ due to neglect of the third term in the right-hand side of Eq. (12) is $(-11.0 \pm 1.8) \times 10^{-10}$ m s$^{-2}$. Clearly, this value compares well with the experimental one of $(-8.7 \pm 0.9) \times 10^{-10}$ m/s$^2$, considering that neither the "Pioneer" experiment nor those forming the basis of this paper were specifically designed to detect such effects. Anomalous accelerations of the same order of magnitude have been also inferred from the motion of Galileo and Ulysses spacecraft [12], but in these two cases the effect could not be well separated from solar radiation pressure effects.

It is useful to note that the above calculations were based on the assumption of constant density of the halo dark matter, which implies that the halo mass varies as $r^3$. The possibility that the mass follows a higher power of $r$, that is $\rho_H$ increases with the distance from Sun, is highly unlikely. On the other hand, for a mass dependency on distance of the type $r^2$ the gravitational force of the Sun's halo would produce a constant acceleration upon *any* test body in the solar system, which would be inconsistent with astrometrical data [12]. It thus follows that a nearly constant $\rho_H$ is a most reasonable choice for the dark matter distribution in the Sun's "neighborhood", as it explains both the value and the constancy of the Pioneer anomalous acceleration.

To conclude this sub-section, it has to be added that the experimental value for the Pioneer 10/11 anomalous acceleration has been calculated [12] as a mean for the two spacecraft, while the present paper proposes that it depends on the spacecraft radial velocity. However, the odds were that the two Pioneers recede in space at almost equal speeds, and their small difference has been therefore included only in the form of speed error in our calculation of the anomalous acceleration. Detailed analysis should be carried out for each spacecraft tracking data separately to better probe their degree of agreement with the predictions of this paper.

### D. Bound orbits and the stability of the Solar System

Upon inspection of Eqs. (10) and (11) one can conclude that the secular mass increase leads to time dependent deviations of the orbital parameters from those corresponding to Keplerian orbits. Such variations can be evaluated by noticing the formal equivalence between our problem and the classical theory for the perturbation of atmospheric drag, $\mathbf{F}/m = -kv\dot{\mathbf{r}}$ [the constant $k$ equals $2\lambda$ in this work – see Eq. (9)], upon the orbits of Earth satellites [32]. That theory typically provides expressions for the variation of orbital parameters, such as the argument of the pericenter $\omega$, semi-major axis $a$, and eccentricity $e$, in terms of integrals over one revolution of some functions of the eccentric anomaly $E$. Accordingly, one gets

$$\delta\omega = -2\lambda a \frac{(1-e^2)^{1/2}}{e} \int_0^{2\pi} \frac{(1+e\cos(E))^{1/2}}{(1-e\cos E)^{1/2}} \sin(E) dE, \tag{14}$$

for the perihelion precession,

$$\delta a = -2\lambda a^2 \int_0^{2\pi} \frac{(1+e\cos(E))^{3/2}}{(1-e\cos E)^{1/2}} dE, \tag{15}$$

for the decay of the semi-major axis, and

$$\delta e = -2\lambda a(1-e^2) \int_0^{2\pi} \frac{(1+e\cos(E))^{1/2}}{(1-e\cos E)^{1/2}} \cos(E) dE, \tag{16}$$

for the eccentricity decay.

Due to the presence of the sin ($E$) under the integral, Eq. (14) gives $\delta\omega = 0$, which proves the statement made in the preceding sub-section that the second term in Eq. (9) does no lead to secular perihelion precession.

In the case of small eccentricities of the unperturbed orbit, Eqs. (15) and (16) can be approximated (up to terms of the second order in $e$) as:

$$\delta a \cong -4\pi\lambda a^2 - 3\pi\lambda a^2 e^2, \tag{17}$$

and:

$$\delta e = -2\pi\lambda ae(1-e^2), \tag{18}$$

respectively. For an object whose unperturbed orbital parameters are the same as those of the Earth, i.e., $a = 1.5 \times 10^{11}$ m and $e = 0.017$, Eq. (17) gives $\delta a = -$ 2200 km/rev, while Eq. (18) leads to $\delta e = -$ $1.2 \times 10^{-7}$ rev$^{-1}$. Such large variations in the semi-major axis have never been observed experimentally for planetary objects. However, significant orbit alterations of non-gravitational origin have been detected for high altitude Earth artificial satellites (e.g., for LAGEOS [33]), though the data available to me permitted only a qualitative evaluation of their agreement with the model proposed here.

In summarizing all the evidence discussed in this paper on the possible existence of a secular mass effect, we are facing the following contrast: both, static (i.e., G measurements between bodies at relative rest) and dynamical measurements (i.e., spacecraft and artificial satellite tracking) suggested that small objects moving in the Sun neighborhood undergo appreciable secular mass increase – a process that has little influence upon the motion of the planets and their natural satellites in order for them to present stable orbits. This calls into question the universality and/or constancy of the parameter $\lambda$ of the present model, which may be negligibly small for planets. Alternatively, one could accept that $\lambda$ is indeed a universal parameter, but that a second process of mass loss (by, e.g., dissipation of internal energy through radiation, or by other mechanism [50]) could intervene to counterbalance the mass increase, in the case of large and/or old objects. For a clear understanding of what the second mechanism might be, one needs to be more specific with regard to the nature of the interaction between dark and baryonic matter and, ultimately, to the nature of the dark matter itself, which is yet another unsolved problem of contemporary physics.

## V. CONCLUSION

My analysis above of the published small-scale gravitational measurements has revealed a surprisingly large secular variation in the gravitational force that, due to the nature of the determinations, appeared formally as a variation in the Newton's gravitational constant G with time. When reinterpreted as a mass increase of the form $\dot{m}/m = \lambda v$ (with λ = constant), caused by impact of the attracting test bodies with the cold dark matter halo of the Sun, this phenomenon predicted substantial deviations from Newtonian dynamics in the solar system, and particularly a cosmic acceleration that for the Pioneer 10/11 spacecraft agrees closely with the published experimental value [12]. (Note that a true secular increase in G at the observed rate, besides contradicting a host of other experiments – see above –, would explain neither the value of the Pioneer acceleration, nor its patent constancy.) For bodies moving on closed orbits, this secular mass effect leads to alteration of their orbital parameters, such as semi-major axis and eccentricity, which is at least qualitatively confirmed by the temporal evolution of the orbits of Earth artificial satellites [33]. By contrast, the effect seems to be negligible for large bodies such as planets and large natural satellites, and this raises the necessity for further studies.

In my search for the best model for the secular mass increase, I have also explored the possibility that $\dot{m}/m = constant$, which would be compatible with a *hot dark matter* scenario, under which even resting bodies may undergo secular mass increase, due this time to rapidly moving dark matter particles impacting the body. I have finally inclined in favor of the cold dark matter scenario since the hot dark matter hypothesis produced a Pioneer acceleration too large by a factor of three compared to the value determined from orbit measurements [12]. However, in the absence of a definitive proof, whether one or the other scenario is correct remains an open question.

It is worth noting that straightforward computations based on known expressions for the gravitational light deflection, redshift, time dilation, and the Shapiro light time delay [5] (taking $M$ as time dependent and assuming that those formulas are exact to the first order in time) show secular variation in these effects that would be beyond the detection limit of any presently known method.

As far as I can see, there are some possibilities for the secular effect discussed in this paper to be independently confirmed. The first hope is that the rate of variation in the gravitational attraction of about $10^{-5}$ year$^{-1}$ obtained in this work should shortly come within the reach of the, e.g., Gundlach et al's apparatus sensitivity [34]. Secondly, detailed analysis should be carried out for the known residuals in the motion of small objects orbiting the Earth or the Sun, which have not been fully explained by conventional physics. The latter method could be more specific, as it might be able to discriminate between $m$ and $G$ variations. A variation in G is already very improbable but, for the meticulous experimentalist, a direct proof could make a difference.

## ACKNOWLEDGEMENTS

Generous support from Prof. T. Sato of the Kochi Medical School is gratefully acknowledged. It is a pleasure to thank Prof. M. Sancho-Ruiz and Prof. A. F. Rañada of the Universidad Complutense de Madrid for very useful comments and criticisms on an earlier version of this work, and also A.S. Cârstea (Institute of Physics and Nuclear Engineering, Bucharest) and D. Niță (Université de Montréal) for useful discussions. Some of the papers cited in this work have been obtained from The NASA Astrophysics Data System (http://adswww.harvard.edu/).

## REFERENCES

[1] T. J. Quinn, Nature **408**, 919 (2000).
[2] P. J. Mohr and B. N. Taylor, Rev. Mod. Phys. **72**, 351 (2000).
[3] E. Fischbach and C. Talmadge, Nature **356**, 207 (1992).
[4] D. R. Long, Phys. Rev. B **9**, 850 (1974).
[5] C. M. Will, Living Rev. Relativity **4**, online article (2001).
[6] C. D. Hoyle, U. Schmidt, B. R. Heckel, E. G. Adelberger, J. H. Gundlach, et al., Phys. Rev. Lett. **86**, 1418 (2001); S. Baeßler, B. R. Heckel, E. G. Adelberger, J. H. Gundlach, U. Schmidt, et al., Phys. Rev. Lett. **83**, 3585 (1999).
[7] J. Binney and S. Tremaine, *Galactic Dynamics* (Princeton University Press, Princeton, 1987).
[8] M. Milgrom, Astrophys. J. **270**, 365 (1983).
[9] R. H. Sanders, Astron. Astrophys. **136**, L21 (1984).
[10] L. Bergström, Rep. Prog. Phys. **63**, 793 (2000).
[11] J. D. Anderson, P. L. Laing, E. L. Lau, A. S. Liu, M. M. Nieto, et al., Phys. Rev. Lett. **81**, 2858 (1998).
[12] J. D. Anderson, P. L. Laing, E. L. Lau, A. S. Liu, M. M. Nieto, et al., gr-qc/0104064 (2001).
[13] M. M. Nieto, J. D. Anderson, P. L. Laing, E. L. Lau, and S. Turyshev, hep-ph/0110373v1 (2001).
[14] W. van Straten, M. Bailes, M. Britton, S. R. Kulkarni, S. B. Anderson, et al., Nature **412**, 158 (2001); T. Damour, Nucl. Phys. B (Proc. Suppl.) **80**, 41 (2000).
[15] J. G. Williams, X. X. Newhall, and J. O. Dickey, Phys. Rev. D **53**, 6730 (1996); J. O. Dickey, P. L. Bender, J. E. Faller, X. X. Newhall, R. L. Ricklefs, et al., Nature **265**, 482 (1994); R. W. Hellings, P. J. Adams, J. D. Anderson, M. S. Keesey, E. L. Lau, et al., Phys. Rev. Lett. **51**, 1609 (1983).
[16] J. H. Taylor, Rev. Mod. Phys. **66**, 711 (1994).
[17] N. Murray, B. Hansen, M. Holman, and S. Tremaine, Science **279**, 69 (1998).
[18] G. T. Gillies, Metrologia **24(Suppl.)**, 1 (1987).
[19] G. T. Gillies, Re. Prog. Phys. **60**, 151 (1997).
[20] R. E. Cohen and B. N. Taylor, J. Phys. Chem. Ref. Data **2**, 663 (1973).
[21] F. D. Stacey, G. J. Tuck, G. I. Moore, S. C. Holding, B. D. Goodwin, et al., Rev. Mod. Phys. **59**, 157 (1987).
[22] O. V. Karagioz and V. P. Izmailov, Mes. Tech. **39**, 979 (1996).
[23] W. Michaelis, H. Haars, and R. Augustin, Metrologia **32**, 267 (1995/96).
[24] T. J. Quinn, C. C. Speake, S. J. Richman, R. S. Davis, and A. Picard, Phys. Rev. Lett. **87**, 111101 (2001).
[25] T. Damour, G. W. Gibbons, and J. H. Taylor, Phys. Rev. Lett. **61**, 1151 (1988).
[26] P. A. M. Dirac, Nature **139**, 323 (1937); P. A. M. Dirac, in *The Physicist's Conception of Nature*, edited by J. Mehra (D. Reidel Publishing Company, Dordrecht, 1973), p. 45.
[27] F. Munyaneza and R. D. Viollier, astro-ph/9910566v1 (1999).
[28] Ø. Grøn and H. H. Soleng, Astrophys. J. **456**, 445 (1996).
[29] J. D. Anderson, E. L. Lau, T. P. Krisher, D. A. Dicus, D. C. Rosenbaum, et al., Astrophys. J. **448**, 885 (1995).
[30] K. Nordtvedt Jr., J. Müller, and M. Soffel, Astron. Astrophys. **293**, L73 (1995).
[31] K. Nordtvedt Jr., Phys. Rev. Lett. **65**, 953 (1990).

[32] A. E. Roy, *The Foundations of Astrodynamics* (The Macmillan Company Inc., New York, 1965).
[33] G. Metris, J. C. Vokrouhlicky, J. C. Ries, and R. J. Eanes, Adv. Space Res. **23**, 721 (1999).
[34] J. H. Gundlach and S. M. Merkowitz, Phys. Rev. Lett. **85**, 2869 (2000).
[35] R. D. Rose, H. M. Parker, R. A. Lowry, A. R. Kuhlthau, and J. W. Beams, Phys. Rev. Lett. **23**, 655 (1969).
[36] G. G. Luther and W. R. Towler, Phys. Rev. Lett. **48**, 121 (1982).
[37] B. Hubler, A. Cornaz, and W. Kündig, Phys. Rev. D **51**, 4005 (1995).
[38] A. Cornaz, B. Hubler, and W. Kündig, Phys. Rev. Lett. **72**, 1152 (1994).
[39] V. P. Izmailov, O. V. Karagioz, V. A. Kuznetsov, V. N. Melnikov, and A. E. Roslyakov, Mes. Tech. **36**, 1065 (1993).
[40] H. Walesch, H. Meyer, H. Piel, and J. Schurr, IEEE Trans. Instr. Meas. **44**, 491 (1995).
[41] M. P. Fitzgerald and T. R. Armstrong, Meas. Sci. Technol. **10**, 439 (1999).
[42] M. P. Fitzgerald and T. R. Armstrong, IEEE Trans. Instr. Meas. **44**, 494 (1995).
[43] C. H. Bagley and G. G. Luther, Phys. Rev. Lett. **78**, 3047 (1997).
[44] J. P. Schwarz, D. S. Robertson, T. M. Niebauer, and J. E. Faller, Science **282**, 2230 (1998); J. P. Schwarz, D. S. Robertson, T. M. Niebauer, and J. E. Faller, Meas. Sci. Technol. **10**, 478 (1999).
[45] F. Nolting, J. Schurr, S. Schlamminger, and W. Kündig, Meas. Sci. Technol. **10**, 487 (1999).
[46] J. Schurr, F. Nolting, and W. Kündig, Phys. Lett. A. **248**, 295 (1998); J. Schurr, F. Nolting, and W. Kündig, Phys. Rev. Lett. **80**, 1142 (1998).
[47] J. Luo, Z.-K. Hu, X.-H. Fu, and S.-H. Fan, Phys. Rev. D **59**, 042001 (1998).
[48] U. Kleinevoß, H. Meyer, A. Schumacher, and S. Hartmann, Meas. Sci. Technol. **10**, 492 (1999).
[49] S. J. Richman, T. J. Quinn, C. C. Speake, and R. S. Davis, Meas. Sci. Technol. **10**, 460 (1999).
[50] In the early stages of this work I have also considered the possibility that for planets the captured mass could take the form of a dark matter halo of gradually increasing size. When the halo becomes large enough, any excess of accreted mass around the planet will be immediately re-captured by the Sun due to its impact with the planet halo or to Sun's stronger gravity. A stationary equilibrium could have been thus attained long time ago, and would have presently little or no effect on the planetary dynamics. However, if this were true, it would be difficult to understand why small objects on the Earth surface do not form external halos as the Earth itself.

**TABLE I. Summary of Newton's constant data published over the past ~75 years.**

| Data Set No. | Author(s) | Date (year) | G ($\times 10^{-11}$ m$^3$ kg$^{-1}$ s$^{-2}$) | Reference |
|---|---|---|---|---|
| 1 | Heyl | 1927 ± 2 | 6.6721 ± 0.0073 | [20] |
| 2 | Heyl & Chrzanowski | 1940.5 ± 0.5 | 6.6720 ± 0.0049 | [20] |
| 3 | Rose et al | 1969.2 ± 0.5 | 6.674 ± 0.003 | [20, 35] |
| 4 | Pontikis | 1971.4 ± 0.1 | 6.67145 ± 0.0001 | [20] |
| 5 | Renner | 1973 ± 1 | 6.670 ±0.008 | [19] |
| 6 | Karagioz et al | 1975 ± 1 | 6.668 ± 0.002 | [19] |
| 7 | Luther et al | 1975 ± 1 | 6.6699 ± 0.0014 | [19] |
| 8 | Sagitov et al | 1978 ± 1 | 6.6745 ± 0.0008 | [19] |
| 9 | Luther & Towler | 1981.5 ± 0.5 | 6.6726 ± 0.0005 | [36] |
| 10 | Karagioz et al | 1985.7 ± 0.3 | 6.6730 ± 0.0005 | [22] |
| 11 | Dousse & Rheme | 1986 ± 1 | 6.6722 ± 0.0051 | [19] |
| 12 | de Boer et al | 1986 ± 1 | 6.667 ± 0.005 | [19] |
| 13 | Karagioz et al | 1986.3 ± 0.1 | 6.6730 ± 0.0003 | [22] |
| 14 | Karagioz et al | 1987.5 ± 0.5 | 6.6730 ± 0.0005 | [22] |
| 15 | Karagioz et al | 1988.7 ± 0.1 | 6.6728 ± 0.0003 | [22] |
| 16 | Karagioz et al | 1989.4 ± 0.1 | 6.6729 ± 0.0002 | [22] |
| 17 | Karagioz et al | 1990.9 ± 0.1 | 6.67300 ± 0.00009 | [22] |
| 18 | Schurr et al | 1991 ± 1 | 6.6613 ± 0.0093 | [19] |
| 19 | Hubler et al | 1992.5 ± 0.4 | 6.6737 ± 0.0051 | [37, 38] |
| 20 | Izmailov et al | 1992.6 ± 0.4 | 6.6771 ± 0.0004 | [39] |
| 21 | Michaelis et al | 1993.2 ± 0.3 | 6.71540 ± 0.00008 | [23] |
| 22 | Hubler et al | 1993.5 ± 0.4 | 6.6698 ± 0.0013 | [37, 38] |
| 23 | Karagioz et al | 1993.6 ± 0.3 | 6.6729 ± 0.0002 | [22] |
| 24 | Walesch et al | 1994.0 ± 0.5 | 6.6719 ± 0.0008 | [40] |
| 25 | Fitzgerald & Armstrong | 1994.0 ± 0.5 | 6.6746 ± 0.001 | [41, 42] |
| 26 | Hubler et al | 1994.3 ± 0.1 | 6.6607 ± 0.0032 | [37] |
| 27 | Hubler et al | 1994.5 ± 0.2 | 6.6779 ± 0.0063 | [37] |
| 28 | Karagioz et al | 1994.8 ± 0.2 | 6.67285 ± 0.00008 | [22] |
| 29 | Karagioz et al | 1995.5 ± 0.3 | 6.6729 ± 0.0002 | [22] |
| 30 | Bagley & Luther | 1996.5 ± 0.5 | 6.6740 ± 0007 | [43] |
| 31 | Schwarz et al | 1997.5 ± 0.1 | 6.686 ± 0.009 | [44] |
| 32 | Schurr, Nolting et al | 1997.5 ± 0.5 | 6.6754 ± 0.0014 | [45, 46] |
| 33 | Luo et al | 1997.8 ± 0.1 | 6.6699 ± 0.0007 | [47] |
| 34 | Schurr, Nolting et al | 1998.3 ± 0.3 | 6.6749 ± 0.0014 | [45] |
| 35 | Schwarz et al | 1998.5 ± 0.1 | 6.689 ± 0.012 | [44] |
| 36 | Kleinevoss et al | 1998.5 ± 0.1 | 6.6735 ± 0.0004 | [48] |

**TABLE I.** ***(Continued.)***

| Data Set No. | Author(s) | Date (year) | G ($\times 10^{-11}$ $m^3$ $kg^{-1}$ $s^{-2}$) | Reference |
|---|---|---|---|---|
| 37 | Richman et al | 1998.5 ± 0.5 | 6.683 ± 0.011 | [49] |
| 38 | Fitzgerald & Armstrong | 1998.7 ± 0.5 | 6.6742 ± 0.0007 | [41] |
| 39 | Gundlach & Merkowitz | 1999.9 ± 0.5 | 6.67422 ± 0.00009 | [34] |
| 40 | Quinn et al | 2000.8 ± 0.5 | 6.67559 ± 0.00027 | [24] |